\documentclass[iop]{emulateapj-rtx4} 
\shortauthors{Sekanina}
\shorttitle{Populations of SOHO Kreutz Sungrazers}
\slugcomment{Version \today}
\newcommand{\Rsun}{$R_{\mbox{\scriptsize \boldmath $\odot$}}$}

\begin{document}
\title{POPULATIONS OF THE KREUTZ SUNGRAZER SYSTEM IN A SOHO DATABASE}
\author{Zdenek Sekanina}
\affil{Jet Propulsion Laboratory, California Institute of Technology,
  4800 Oak Grove Drive, Pasadena, CA 91109, U.S.A.}
\email{Zdenek.Sekanina@jpl.nasa.gov.}

\begin{abstract} 
%
Discovery of nine populations in a set of 193 select SOHO Kreutz sungrazers         
(Sekanina 2021) is confirmed for the first time via a histogram of the {\it true\/} 
longitudes of the ascending node, constructed for a revised set of 220~select       
sungrazers imaged exclusively by the SOHO's C2 coronagraph.  Marsden's orbits       
are approximately corrected for effects of the out-of-plane  nongravitational       
force.  Population~I displays two peaks in the histogram, one presumably belonging  
to a side branch alike to Population~Pe, but with no related naked-eye sungrazer    
known.  Swarms/clusters of objects are commonplace, providing evidence on           
cascading fragmentation proceeding throughout the orbit.  Augmentation to {\it      
all\/} C2-only SOHO Kreutz comets, aimed at removing deliberate bias against        
Populations~I and Pe, reduces the appearance of Populations~Ia and \mbox{Pre-I}     
to bulges along the slopes of the histogram because of the swollen wings of         
Populations~I and Pe, respectively.  Populations~II through IV change very          
little or not at all.  The high Population~I-to-II abundance ratio, of 14:1,        
may be a product~of~temporal limitations in fragment release.  A drop in            
the number of fragments toward the ends~of~the~nodal-longitude distribution,        
especially from Population~II to IV, is in line with the contact-binary model.      
\end{abstract} 
\keywords{comets general: SOHO sungrazers; comets individual: X/1106 C1, C/1843 D1,
 C/1882\,R1, C/196\,R1, C/1965\,S1, C/1970\,K1, C/2011\,W3; methods: data
 analysis{\vspace{-0.1cm}}}
\section{Introduction}  
More than 4500 comets had been discovered by the end of October 2022 in the images
taken by the C2 and C3 LASCO coronagraphs on board the {\it Solar and Heliospheric
Observatory\/} (SOHO), of which an estimated 85~percent are members of the Kreutz
sungrazer system.  With a little over 200~comets typically reported per year (e.g.,
Battams \& Knight 2017), the corresponding average discovery rate of the Kreutz
sungrazers by SOHO is almost exactly one per two days, even though strongly season
dependent.

It has long been known that orbitally the Kreutz system is extremely complex
(e.g., Kreutz 1901, Marsden 2005).  Marsden (1967) and others remarked on two
major subgroups, one represented by the Great March Comet of 1843 (C/1843~D1),
the other by the Great September Comet of 1882 (C/1882~R1) and comet Ikeya-Seki
(C/1965~S1).  The subgroups' orbits differ in both the perihelion distance and
angular elements, the largest disparity of nearly 20$^\circ$ in the longitude of the
ascending node.  Three years after Marsden's paper was published, this classification
was defied by comet White-Ortiz-Bolelli (C/1970~K1), necessitating the introduction
of a third subgroup (Marsden 1989, 1990).  The extended model held until 2011, when
the orbit of a new sungrazing comet Lovejoy (C/2011~W3) made it indispensable to add
a fourth population (to use the term I prefer to subgroup), as argued for example by
Sekanina \& Chodas (2012).  A novel method, recently used to examine Marsden's orbits
for 193 select objects, a subset of 1565~SOHO Kreutz sungrazers from 1996--2010,
revealed a total of {\it nine populations\/} (Sekanina 2021; referred to hereafter
as Paper~1).  The present study displays the populations in a different manner
for the first time, augmenting their distribution by employing a greatly expanded
dataset.

\section{New Portrayal of the SOHO Kreutz Populations} 
All bright, naked-eye Kreutz sungrazers share a common line of apsides.  On this
condition it was alarming to see Marsden's orbits of the SOHO Kreutz sungrazers to
exhibit a strong dependence of the latitude of perihelion on the longitude of the
ascending node, by up to $\sim$25$^\circ$ over a nodal-longitude range of nearly
90$^\circ$.  This anomaly was shown by Sekanina \& Kracht (2015) to be a product
of the neglected effects of the {\it out-of-plane\/} component of the
sublimation-driven nongravitational force.

Forcing the standard orientation of the line of apsides effectively circumvents the
need to compute the magnitude of the acceleration, replacing a {\it nominal\/} value
of the longitude of the ascending node, $\Omega$, in {\vspace{-0.08cm}}Marsden's orbit
with its {\it true\/} value, $\widehat{\Omega}$.  The difference{\vspace{-0.085cm}}
between the nominal and true nodal longitudes, \mbox{$|\Omega - \widehat{\Omega}|$},
was shown by Sekanina \& Kracht (2015) to be a crude proxy for the parameter of the
out-of-plane nongravitational effect to which a given SOHO sungrazer was subjected:\
the greater the difference, the higher the parameter's value.

In a plot of the set of 193 select SOHO Kreutz objects, whose orbits were determined
exclusively from the astrometric positions of the C2 coronagraphic images, the nominal
latitude of perihelion, $B_\pi$, was shown in Paper~1 to satisfy with high accuracy
a linear relationship with the nominal longitude of the ascending node, $\Omega$,
\begin{equation}
B_\pi = \widehat{B}_\pi + b \left( \Omega - \widehat{\Omega} \right),  
\vspace{-0.17cm}
\end{equation}
where $\widehat{B}_\pi$ is the standard latitude of perihelion of the naked-eye Kreutz
sungrazers, $b$ is a dimensionless slope coefficient, \mbox{$b = 0.28$}, and
\begin{equation}
B_\pi = \arcsin ( \sin \omega \sin i ). 
\end{equation}

\begin{figure*}[t]
\vspace{-11.3cm}
\hspace{1.4cm}
\centerline{
\scalebox{0.86}{
\includegraphics{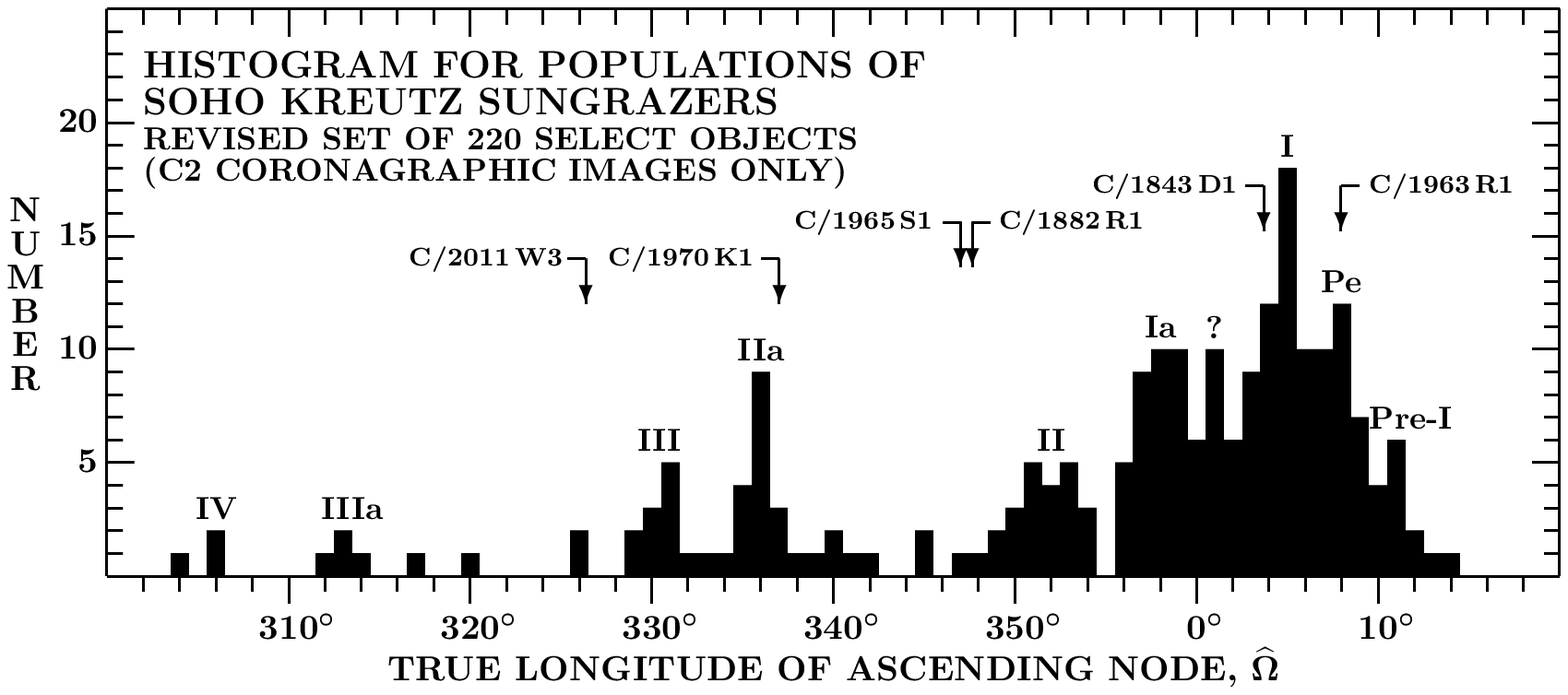}}} 
\vspace{-8cm}
\caption{Histogram of the true longitudes of the ascending node for the revised set of
 220~select SOHO Kreutz sungrazers, whose orbits were derived exclusively from the
 C2 coronagraphic images.  The nine populations are marked.  Population~Pe is considered
 a side branch of Population~I.  Also depicted are the nodal longitudes of six
 associated naked-eye sungrazers.  With the exception of Populations~Pe and IIa the
 peaks of the related SOHO sungrazer swarms or clusters appear to be shifted by a few
 degrees toward the larger longitudes.  An additional, previously undetected cluster has
 a peak at the nodal longitude of 1$^\circ$, labeled by a question mark and probably
 another side branch of Population~I with no known associated naked-eye
 object.{\vspace{0.61cm}}}
\end{figure*}

The nominal latitude of perihelion is here expressed as a function of the nominal
argument of perihelion, $\omega$, and nominal inclination, $i$.  The true longitude
of the ascending node of a $k$-th SOHO object is then given~by~an~expression
\begin{equation}
\widehat{\Omega}_k = \Omega_k + \frac{1}{b} \left[ \widehat{B}_\pi - (B_\pi)_k \right].
\end{equation}

The nine populations of the SOHO sungrazers were detected in Paper~1 as sets of data
points distributed along nine parallel lines in the plot of the nominal latitude
of perihelion against the nominal nodal longitude, \mbox{$B_\pi = B_\pi(\Omega)$},
shifted relative to each other in the ordinate.  On each such line, the abscissa of
the point at which the nominal and{\vspace{-0.08cm}} true nodal longitudes coincide
marks the {\it averaged\/} true value of the nodal longitude $\langle \widehat{\Omega}
\rangle$ of the population, which is given by
\begin{equation}
\langle \widehat{\Omega} \rangle = \frac{1}{\nu} \sum_{k=1}^{\nu} \widehat{\Omega}_k,
\end{equation}
where $\nu$ is the number of the sungrazers in the population.

In the mentioned plot the range of the population's nominal nodal longitudes (i.e.,
the scatter {\it along\/} the line) is --- as already briefly noted --- a measure
of the overall span of the out-of-plane nongravitational accelerations that the
plotted sungrazers were subjected to, whereas the minor scatter of the points in
the ordinate (i.e., the scatter {\it across\/} the line) provides information on
the effects of progressive fragmentation of the SOHO sungrazers.  A tendency of two
neighboring populations to overlap suggests that the overall range of fragmentation
scatter in the true longitude of the ascending node is comparable to the size of
the gap between the nodal longitudes of the populations.  

As determined in Paper~1, in the order of decreasing true nodal longitude the detected
populations were \mbox{Pre-I}, Pe, I, Ia, II, IIa (with a sub-population IIa*), III,
IIIa, and IV.  Population Pe, associated with comet Pereyra (C/1963~R1), is considered
a side branch of Population~I.   The set of the 193 select SOHO Kreutz sungrazers,
which was assembled in Paper~1 and tabulated in a recent summary paper (Seka\-nina
2022; referred to hereafter as Paper~2), was markedly biased against Populations~I,
Pe, and \mbox{Pre-I}, because included were only the sungrazers, for which Marsden's
gravitational orbits were based on at least 12~astrometric observations.  Less
strongly biased was Population~Ia, for which the minimum required number of
astrometric observations was 10.  Other populations had the lower limit of only
five observations.

By performing the operation (3) it is possible to portray the populations in a
manner very different from the parallel straight lines, because any SOHO
sungrazer is now described not by two{\vspace{-0.09cm}} quantities --- $\Omega$
and $B_\pi$ ---~but~by a single quantity, the true nodal longitude
$\widehat{\Omega}$.  The obvious aim is to examine the distribution
of the true nodal longitudes by generating their {\it histogram\/}.

Re-inspection of Marsden's catalogue of 1500+ gravitational orbits for the SOHO
Kreutz sungrazers showed the need to revise the set of the 193~comets imaged
exclusively by the C2 coronagraph on board the SOHO spacecraft.  As a result, a new
set of 220~select objects, displayed in Figure~1, was obtained by removing a few
entries (whose astrometry turned out to be contaminated by imaging with the C3
coronagraph) from the existing set and adding 30~entries that previously were
missed.  Several entries needed to be reclassified, but they had of course no
effect on the distribution of the longitudes of the ascending node.  The deliberate
bias against Populations~I and Pe, introduced in the original set to mitigate the
dominant influence of Population~I (and its side branch, Population~Pe), was
preserved by allowing to include only the members whose orbits were computed
from at least 12~C2 astrometric positions.  The bias was slightly reduced for
Population~\mbox{Pre-I}, which now included members with the orbits based on at
least 11~positions.  The lower limits for the other populations were not changed:\
10~positions for Population~Ia and five for{\nopagebreak} \mbox{Populations~II}
through IV.{\pagebreak}
\begin{table}[t]
\vspace{-4.11cm}
\hspace{5.25cm}
\centerline{
\scalebox{1}{
\includegraphics{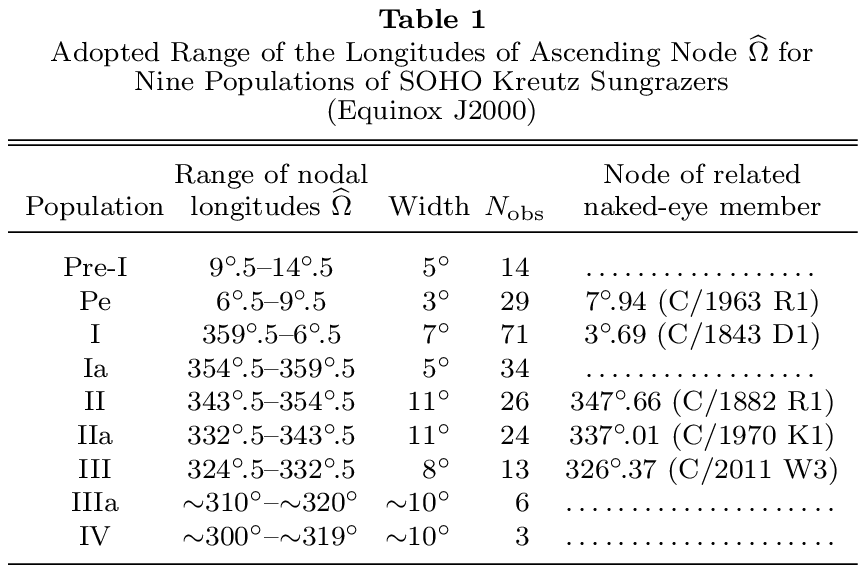}}} 
\vspace{-19.25cm}
\end{table}

Although the line of apsides is considered to be fixed, it actually differs slightly
from population to population.  Although neglecting this effect influences the true
nodal longitudes by only a small fraction of 1$^\circ$, I did take it into account
by using a standard latitude of perihelion equaling +35$^\circ\!$.29 for Population~I
and +35$^\circ\!$.26 for Population~Ia.  For Populations~II through IV I adopted
a value of +35$^\circ\!$.23 and for Populations~Pe and \mbox{Pre-I} a value of
+35$^\circ\!$.33 (all for Equinox J2000).

With an adopted step of 1$^\circ$, the histogram for the set of the 220~select
SOHO Kreutz sungrazers in Figure~1 largely confirms the introduction in
Paper~1 of nine detected populations.  There are however two caveats.  One,
an additional peak, at a nodal longitude of 1$^\circ$, is located between
the peaks of Populations~I and Ia.  This is probably another side branch of
Population~I, with an associated brilliant, naked-eye sungrazer still awaiting
discovery.  And two, except for Populations~Pe and IIa, the peaks of the SOHO
sungrazer clusters appear to be moved by a few degrees toward the greater
true longitudes of the ascending node than are the nodal longitudes of the
related bright Kreutz sungrazers.  The respective shifts are $\sim$1$^\circ$
for Population~I, $\sim$3$^\circ$ to $\sim$5$^\circ$ for Population~II, and
$\sim$4$^\circ$ for Population~III.

Keeping the caveats in mind, I note that in general no gaps are evident
between the neighboring populations (except when the number of members is
low).  This is not surprising, as overlaps can be due to two obvious,
though very different, causes of data scatter.  One is dynamical in nature:\
given that all SOHO sungrazers are fragments, and a significant fraction
presumably high-generation fragments, the scatter is a cumulative product of
the orbital perturbations associated with multiple breakup events.  The other
cause is due to errors in measuring and reducing the astrometric positions,
which are responsible for low accuracy of Marsden's orbital elements.

The overlaps of populations notwithstanding, it is useful to estimate a range of
the (true) longitudes of the ascending node over which each population dominates.
The other two angular elements are determined by the nodal longitude and the fixed
position of the line of apsides.  The perihelion distance and the perihelion time
are besides the nodal longitude the only other independent elements of the
parabolic orbits.

Table 1 presents the range of the longitudes of the ascending node that I adopted for
each of the nine populations, the wings of some extending across the postulated
limits of the assigned spans as noted above.  If a naked-eye sungrazer is related
to the population in that its nodal longitude is in the same interval, it is, with
its osculating value of the longitude, presented in column~5; if more such objects
are known, the intrinsically brightest one is listed.  The limits of Populations~IIIa
and IV are poorly defined because of extremely low numbers of members, as seen from
Figure~1.  

The feature to the immediate left of the main peak of Population~I in Figure~1 is,
unlike in the case of Pe, not classified as a separate population, primarily because
no naked-eye sungrazer with a nodal longitude near 1$^\circ$ is known.  On the other
hand, it is appropriate to again emphasize the bias:\ the memberships of Populations~I
and Pe --- and to a lesser degree, Populations~\mbox{Pre-I} and Ia --- are in Figure~1
underestimated in comparison with the memberships of Populations~II through
IV.  This underscores the enormity of the difference between the total number of
fragments in the SOHO Populations~I and II.  The major deficit of minor members
in Population~II is either related to dramatic differences in the fragmentation
properties of the two populations or there is a severe selection effect.  I will
return to this issue in Section~5 of this paper.

\begin{table}[t]
\vspace{-4.11cm}
\hspace{5.3cm}
\centerline{
\scalebox{1}{
\includegraphics{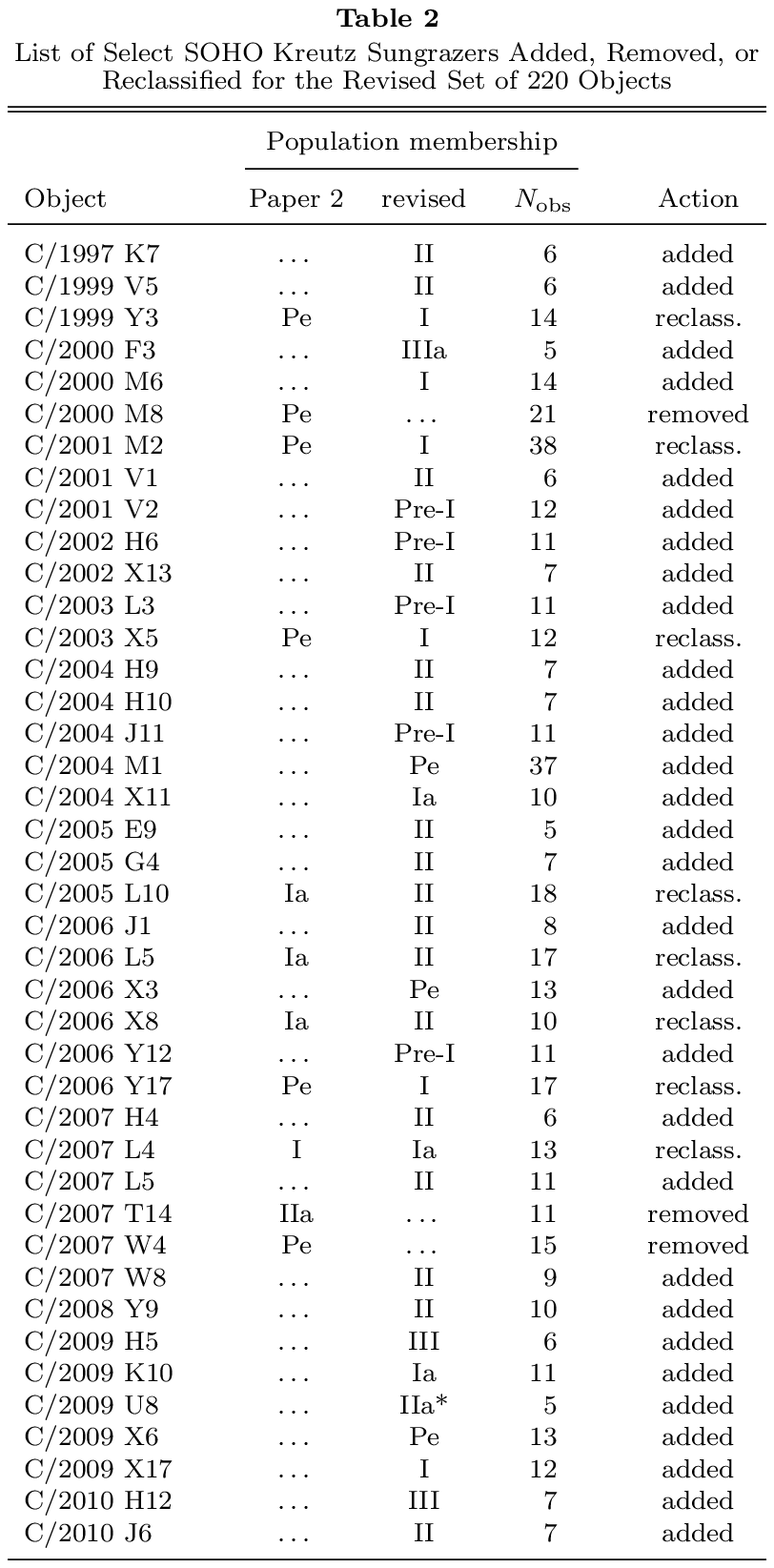}}} 
\vspace{-8.4cm}
\end{table}

The adoption of a specific range of the nodal longitudes for each population
provided the basis for revising its membership in the set of select SOHO
sungrazers.  Together with the Kreutz sungrazers added to, or removed from, the
original set of 193~select objects in the course of re-inspection of Marsden's
catalogue of the SOHO comets, several objects had to be reclassified in order to
fit the correct populations in Table~1.  The 41 cases of additions, eliminations,
and reclassifications are summarized in Table~2.  The search technique employed
in the process of re-inspection as well as the encountered pitfalls are described
in Section~4.  Here I only remark that for dates before mid-March 2005 only a
prohibitively time-consuming method would in {\it every single case\/} reliably
discriminate between the objects imaged exclusively with the C2 coronagraph, which
were of interest to this investigation, and the objects imaged either exclusively
with the C3 coronagraph or with both coronagraphs.

\section{Pairs and Clusters of SOHO Kreutz Sungrazers As Fragments} 
Marsden (1989) commented on two pairs of Kreutz sungrazes detected by a coronagraph
on board the {\it Solar Maximum Mission\/} (SMM) that approached perihelion in
close succession, less than two weeks apart.  The coronagraphs on board SOHO have
routinely been detecting pairs and less often clusters of Kreutz sungrazers on
timescales as short as a fraction of a day.  These objects were proposed to be
the outcome of a sequence of events of nontidal cascading fragmentation, which
--- following a tidal breakup at perihelion --- have continued along the outbound
leg of the orbit to and, presumably, past aphelion (Sekanina 2000).

It turns out that the nearly-simultaneous arrival of the SOHO sungrazers in pairs
and/or clusters is not the only signature of their strongly nonuniform orbital
distribution.  As an example, Table~3 lists eight SOHO sungrazers from the set
plotted in Figure~1, whose true longitudes of the ascending node equaled almost
exactly 4$^\circ\!$.5 and were thus very close to both the main peak of
Population~I and the nodal line of the spectacular 1843 sungrazer.  For each
of the eight sungrazers I tabulate the difference between the nominal and true
nodal longitudes,
\begin{equation}
\mbox{$\partial \Omega = \Omega - \widehat{\Omega}$}, 
\end{equation}
which shows that the overall true-longitude range of 0$^\circ\!$.2 is equivalent
to an overall range of nominal longitudes of nearly 22$^\circ$!  As noted in
Section~2, $\partial \Omega$ is an approximate measure of the cumulative effect
of the sublimation-driven nongravitational acceleration in the direction normal
to the orbital plane.  A set of $n$ objects is characterized by an averaged
quadratic difference \mbox{$\langle \partial \Omega \rangle$} given by
\begin{equation}
\langle \partial \Omega \rangle = \sqrt{\frac{1}{n} \sum_{k=1}^{n}
 \left( \partial \Omega_k \right)^{2}}. 
\end{equation}
For the eight sungrazers in Table 3 I find \mbox{$\langle \partial \Omega
\rangle = 6^\circ\!$.72}.  Judging from an approximate correlation between
the nongravitational parameter in the out-of-plane direction, $A_3$, and the
magnitude of the effect in the nodal longitude in Table~4 of Sekanina \& Kracht
(2015), this value of $\langle \partial \Omega \rangle$ implies{\vspace{-0.03cm}}
an approximate average of \mbox{$\langle A_3 \rangle \simeq 10^{-5}$\,AU
day$^{-2}$}, about three orders of magnitude higher than is expected for
the {\it radial\/} component of the nongravitational acceleration exerted
on a cometary nucleus a kilometer or so across and $\sim$30~times higher than
the long-period comet C/1998~P1 with an anomalously large nongravitational
effect (Marsden \& Williams 2008).  

The remaining columns of Table 3 list the nominal perihelion distance, $q$, and
perihelion time, $t_\pi$; the overall time span is a few weeks short of 12~years.
The nominal perihelion distance cannot readily be converted into a true perihelion
distance; the only general conclusion from the extensive computations by Sekanina
\& Kracht (2015) on a few SOHO Kreutz sungrazers is that the true perihelion
distances have a tendency to be lower than the nominal values.  The perihelion
time is largely immune to the nongravitational forces because they are limited
to small heliocentric distances at which the process of fragmentation is, if in
progress at all, in its early stage and the fragments' dimensions are too large
to be subjected to substantial sublimation-driven effects.
\begin{table}[b]
\vspace{0.8cm} 
\hspace{-0.19cm}   
\centerline{
\scalebox{1}{
\includegraphics{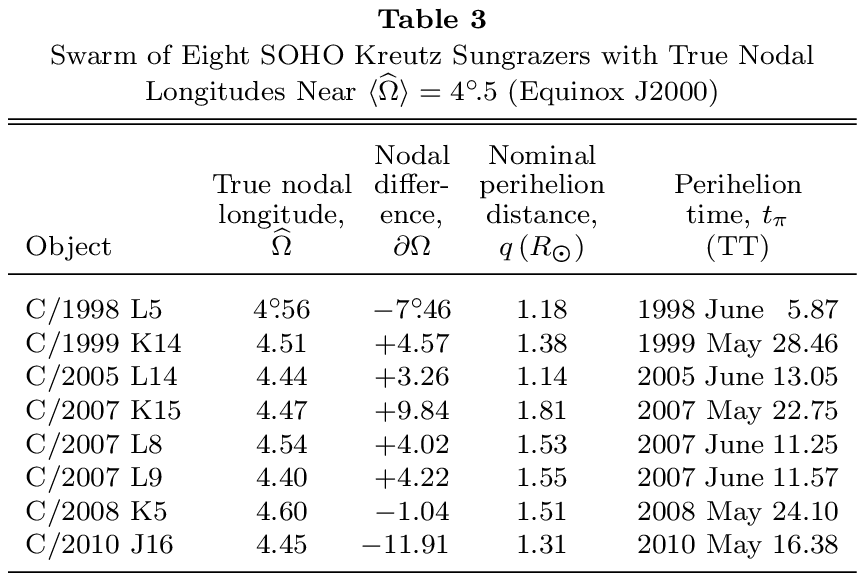}}} 
\end{table}

Table 3 demonstrates that a coincidence of true longitudes of the ascending node
is another property of SOHO Kreutz sungrazers that corroborates the argument on
their highly nonuniform orbital distribution.  Among the tabulated objects is
a pair arriving almost simultaneously, within 8~hr of one another, and a third
object less than 3~weeks earlier, reminiscent of Marsden's example of the SMM
sungrazers.

\begin{table*}[t]
\vspace{-4.1cm}
\hspace{0.54cm}
\centerline{
\scalebox{1}{
\includegraphics{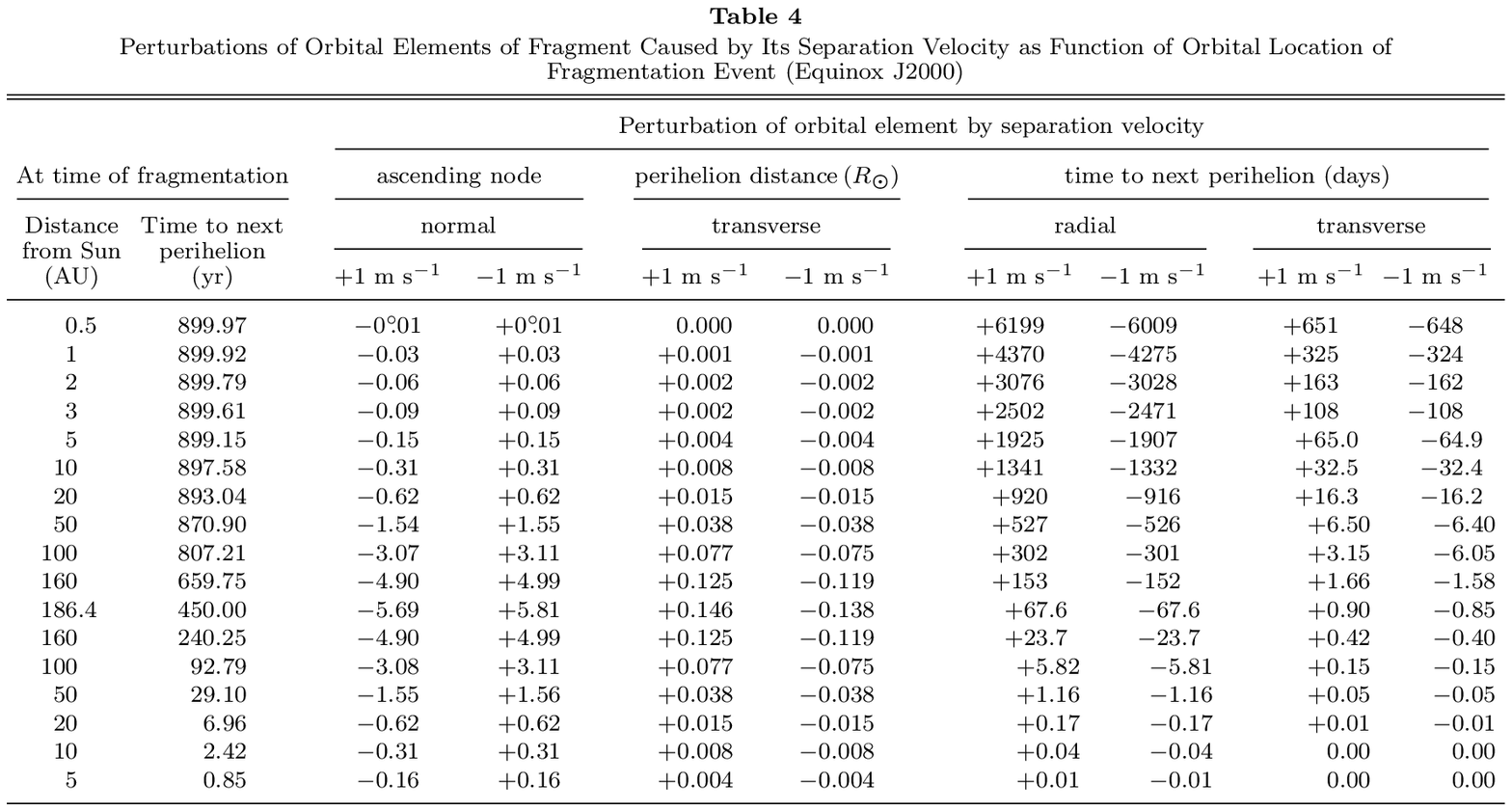}}} 
\vspace{-15.28cm}
\end{table*}

While all or nearly all of the eight objects in Table~3 are obviously rather closely
related to one another, it is not possible to pinpoint the sequence of fragmentation
events responsible for their birth.  However, some insight is gained by comparing the
{\it averaged\/} scatter among the eight objects in the orbital elements.  Let the true
longitude of the ascending node{\vspace{-0.065cm}} of a $k$-th tabulated sungrazer be
$\widehat{\Omega}_k$, its nominal perihelion distance $q_k$, and its perihelion time
$(t_\pi)_k$; the {\it averaged\/} quadratic differences between any two among the $n$
objects are{\vspace{-0.25cm}}

\begin{eqnarray}
\langle \Delta \widehat{\Omega} \rangle & = & \sqrt{\frac{2}{n(n-1)} \sum_{(i \neq j)} 
 \left( \!\widehat{\Omega}_j - \widehat{\Omega}_i \! \right)^{\!2}}, \nonumber \\[0.1cm]
\langle \Delta q \rangle & = & \sqrt{\frac{2}{n(n-1)} \sum_{(i \neq j)} \left(
 q_j - q_i  \right)^{2}}, \nonumber \\[0.1cm]
\langle \Delta t_\pi \rangle & = & \sqrt{\frac{2}{n(n-1)} \sum_{(i \neq j)} \left[
 (t_\pi)_j - (t_\pi)_i \right]^{2}}. 
\end{eqnarray}
Applying (7) to the eight objects in Table 3, I get
\begin{eqnarray}
\langle \Delta \widehat{\Omega} \rangle & \, = \, & 0^\circ\!.096, \nonumber \\[0.05cm]
\langle \Delta q \rangle & \, = \, & 0.31 \, \mbox{\Rsun}, \nonumber \\[0.05cm]
\langle \Delta t_\pi \rangle & \, = \, & 2229 \; \mbox{days}.{\vspace{0.1cm}} 
\end{eqnarray}

In order to find out what do these numbers mean in terms of ``effective'' orbital
locations of fragmentation events as a function of the separation velocity, the
computed perturbations of the three elements in Equations (7) and (8) extracted from
more extensive tables are presented in Table~4.  It shows that the effects on the
longitude of the ascending node are triggered by the separation velocity's normal
(out-of-plane) component, the effects on the perihelion distance by the transverse
component, and the effects on the perihelion time primarily by the radial component,
with a minor contribution from the transverse component.  As the objects in Table~3
belong to Population~I, the computations were made for the orbit of the Great March
Comet of 1843.  In line with my preliminary model for the SOHO Kreutz sungrazers
in Paper~2 (see its Section 4.6), I assume that the fragments split from the Great
Comet of 1106 at perihelion, requiring an orbital period of just about 900~years to
return at the beginning of the 21st century.  Table~4 is computed for a velocity of
$\pm$1~m~s$^{-1}$ in each component in the right-handed RTN orthogonal coordinate
system, with the positive radial direction away from the Sun and the positive normal
direction toward the northern orbital pole.  The orbital effects vary approximately
linearly with the separation velocity as long as it remains low.

The table illustrates quantitatively what was said in Papers~1 and 2 in more general
terms.  In particular, the perturbations --- per 1~m~s$^{-1}$ separation velocity ---
of the nodal longitude and perihelion distance peak, at nearly 6$^\circ$(!) and
0.15~{\Rsun}, respectively, when the fragmentation event occurs at aphelion.  By
contrast, the time of the next perihelion passage is perturbed most when the event
takes place at or near perihelion, the magnitude of the perturbation declining
systematically with time.

Comparison of the effective values from (8) with the tabulated values provides some
interesting results.  First of all, the {\it nominal\/} perihelion distances are
meaningless, offering a perturbation that is much too large by a wide margin.  On
the other hand, both the nodal longitudes and the perihelion times do, rather
surprisingly, correspond to approximately the same heliocentric distance at
fragmentation when the separation velocity is near 1~m~s$^{-1}$.  Assuming a
single event and neglecting the contribution from the transverse component, the
eight sungrazers in Table~3 could derive from a parent that fragmented at 3~AU
(0.4~yr after the 1106 perihelion) with an effective radial separation velocity
of $\sim$0.9~m~s$^{-1}$ and an effective normal separation velocity of
$\sim$1.1~m~s$^{-1}$.  If the parent fragmented instead at 2~AU from the Sun
(0.2~yr after the 1106 perihelion), the radial velocity should be $\sim$0.7~m~s$^{-1}$
and the normal velocity $\sim$1.6~m~s$^{-1}$.  If at 5~AU (0.85~yr after perihelion),
the numbers are $\sim$1.2~m~s$^{-1}$ and $\sim$0.6~m~s$^{-1}$, respectively.  Because
the perturbation of the nodal longitude increases with time, while the perturbation of
the time of next perihelion decreases with time, the range of plausible fragmentation
times is limited, if the radial and normal velocities should be of comparable
magnitudes.  One also notes that a second solution that fits the
perturbation of the nodal longitude, a fraction of a year prior to observation,
is unacceptable because it requires that the eight sungrazers of Table~3
arrive essentially simultaneously, within a few days of each other.

\begin{table}[b]
\vspace{0.8cm} 
\hspace{-0.19cm}  
\centerline{
\scalebox{1}{
\includegraphics{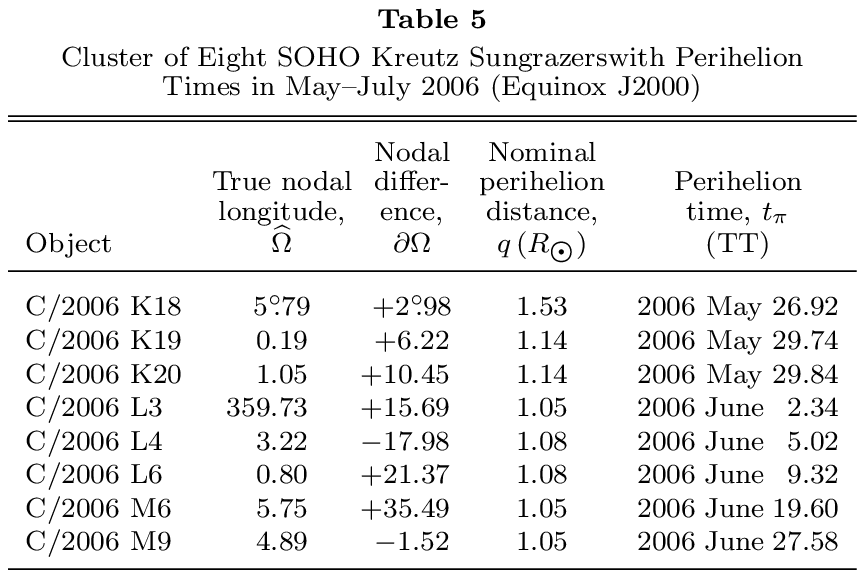}}} 
\end{table}

The assumption of a single fragmentation event is unrealistic because it is in
conflict with the standing hypothesis of {\it cascading fragmentation\/}.  Let each
of the eight sungrazers in Table~3 be a product of a sequence of $n$ fragmentation
events and let in each step of the process the parent fragment split into two
subfragments of equal characteristic dimension.  If the characteristic dimension
of a tabulated{\vspace{-0.08cm}} sungrazer is ${\cal D}_n$, the characteristic
dimension of its direct parent is \mbox{${\cal D}_{n-1} = 2^{\frac{1}{3}} {\cal D}_n$},
etc.  The characteristic dimension{\vspace{-0.08cm}} of the parent of a $k$-th
generation is \mbox{${\cal D}_k = 2^{\frac{1}{3}}{\cal D}_{k+1} = 2^{\frac{n-k}{3}}
{\cal D}_n$} and for the original{\vspace{-0.05cm}} parent, ${\cal D}_0$, it obviously
equals \mbox{${\cal D}_0 = 2^{\frac{n}{3}} {\cal D}_n$}.  The distance between the
centers of mass of two fragments of a $k$-th generation (\mbox{$k \leq n$}) can
be shown to become \mbox{$\zeta {\cal D}_{k-1}$}, where statistically \mbox{$\zeta =
0.3473\ldots$}\,.  If $P_{\rm rot}$ is the rotation period, the two fragments move
away from each other with a separation velocity equaling twice the rotation velocity
of their centers of mass or \mbox{$2 \pi \zeta {\cal D}_{k-1}/P_{\rm rot}$}.  The
perturbations of the orbital elements of a sungrazer in Table~3 (as a fragment of
an $n$-th generation) reflect the integrated effect of the perturbations of all its
parent fragments, caused by the sum $V_{\rm sum}$ of the separation velocities
\begin{equation}
V_{\rm sum} = \frac{2 \pi \zeta}{P_{\rm rot}} \sum_{k=0}^{n-1} {\cal D}_k. 
\end{equation}
This is of course the case of a hypothetical, maximum effect when the
perturbations accumulate in the same direction.  More realistically,
the total perturbation is a square root of the sum of squares of the
perturbations of the fragments of the individual generations, so that,
using in part the properties of the geometric series,
\begin{eqnarray}
V_{\rm sum} & = & \frac{2 \pi \zeta}{P_{\rm rot}} \sqrt{\sum_{k=0}^{n-1} {\cal D}_k^2}
                  \nonumber \\[0.1cm]
            & = & \frac{2 \pi \zeta}{P_{\rm rot}} {\cal D}_0
                  \sqrt{\sum_{k=0}^{n-1} 2^{-\frac{2k}{3}}} \nonumber \\[0.1cm]
            & \simeq & \frac{2 \pi \zeta}{P_{\rm rot}} {\cal D}_0
                  \sqrt{\sum_{k=0}^{\infty} 2^{-\frac{2k}{3}}} \nonumber \\[0.1cm]
            & = & \frac{2 \pi \zeta}{P_{\rm rot}} \frac{{\cal D}_0}{\sqrt{1 -
                   2^{-\frac{2}{3}}}} \nonumber \\[0.1cm]
            & \simeq & \frac{3.59 {\cal D}_0}{P_{\rm rot}}. 
\end{eqnarray}
The approximation used employs the fact that the contributions from the
generations of ever smaller fragments are also progressively smaller.
This is not only logical but is supported by the fact that a very
small fragment would separate from the original parent with a
velocity of \mbox{$\pi {\cal D}_0/P_{\rm rot}$}, which happens to amount
to 7/8 the velocity in (10).  The hypothesis of a single fragmentation
event is therefore not really useless in the light of this finding, at least
as a first approximation.  The likelihood that the fragments spin up as
their dimensions diminish could of course increase the effect of multiple
fragmentation to a degree.  Because of the wide range of perihelion times,
the data are also affected to some extent by the differential planetary
perturbations.

The results for the swarm of SOHO sungrazers in Table~3 should not be
understood to mean that fragmentation is only limited to early months or,
at most, a few years after the parent comet's initial tidal breakup near
perihelion.  The results depend of course on the choice of swarm
sungrazers; in Table~3 the selection was dictated by a common value of
the nodal longitude.  A tight cluster of eight sungrazers from the set of
220 select SOHO comets, also belonging to Population~I, is presented in
Table~5.  Because they arrived at perihelion --- unlike the objects in
Table~3 --- at about the same time, within two weeks or so of mid-June
2006, the differential planetary perturbations were trivial.  The
computation of their averaged quadratic differences gave
\begin{eqnarray}
\langle \Delta \widehat{\Omega} \rangle & \, = \, & 3^\circ\!.60,
   \nonumber \\[0.05cm]
\langle \Delta q \rangle & \, = \, & 0.23 \, \mbox{\Rsun}, \nonumber \\[0.05cm]
\langle \Delta t_\pi \rangle & \, = \, & 15.77 \; {\rm days}. 
\end{eqnarray}  
Comparison with the data{\vspace{-0.05cm}} in Table 4 shows that $\langle \Delta
q \rangle$ is again much too high and useless, while $\langle \Delta \widehat{\Omega}
\rangle$ and $\langle \Delta t_\pi \rangle$ imply an effective fragmentation
time at large heliocentric distance {\it post-aphelion\/}, if the normal
and radial separation velocities were comparable in magnitude:\ the normal
was 0.7~m~s$^{-1}$ at 160~AU from the Sun, 0.9~m~s$^{-1}$ at 130~AU, and
1.2~m~s$^{-1}$ at 100~AU; and the radial one was 0.7~m~s$^{-1}$ at 160~AU,
1.3~m~s$^{-1}$ at 130~AU, and 2.7~m~s$^{-1}$ at 100~AU.  At aphelion the
required values of the two velocity components would be 0.6 and 0.2~m~s$^{-1}$,
respectively.  The normal velocity could not be lower than 0.6~m~s$^{-1}$.

A third example that I consider is a pair of relatively bright SOHO Kreutz
sungrazers, C/1998~K10 and C/1998~K11, whose perihelion times were 0.18~day
apart and were clearly fragments of a larger object.  They both reached a
peak magnitude of about 0.  Even though they did not survive the perihelion
passage, their substantial brightness suggested that their dimensions
significantly exceeded the dimensions of most other SOHO sungrazers (including
those in Tables~3 and 5).  The two 1998~objects are not listed among the
220~select SOHO sungrazers because they were heavily imaged by the C3 coronagraph.
They are members of Population~I, their true longitudes of the ascending
node differ by 5$^\circ\!$.0, and their nominal perihelion distances by
0.17\,{\Rsun}.  There is no question that they did not follow each other in
the same path; comparison of the difference in the nodal longitude with
the computations in Table~4 suggests that the normal separation velocity
could not be lower than 0.9~m~s$^{-1}$, even if the two sungrazers separated
from their common parent at aphelion.

The difference of 4~hours between the two 1998 sungrazers in their
perihelion times offers a very different result.  It requires a radial
separation velocity of 1~m~s$^{-1}$ at a fragmentation time of $\sim$22~AU
post-aphelion, but at aphelion the needed radial velocity is only
{\vspace{-0.04cm}}0.003~m~s$^{-1}$\,(!) or a transverse velocity of
0.2~m~s$^{-1}$.  It appears that the normal component dominates.  The
same conclusion was reached by Sekanina \& Kracht (2015) about the
sublimation-driven nongravitational acceleration affecting the motions of
the SOHO sungrazers near the end of the trajectory.  However, no
nongravitational acceleration could play a role at extremely large heliocentric
distance, where sublimation (especially of less volatile substances) is
out of question.

It is highly probable that fragmentation of the Kreutz sungrazers among the
SOHO objects continues episodically throughout the orbit, resulting in a
systematic size-distribution effects.  The fragments begin as relatively large,
subkilometer to kilometers-sized bodies, many of which keep splitting into
ever smaller pieces.  Others suffer relatively little disruption, arriving
at perihelion as more sizable objects.  The separation velocities,
submeter-per-second to meter-per-second, inferred to fit pairs and clusters
or swarms of SOHO Kreutz sungrazers, are of unclear origin.  If they are
rotational in origin, many fragments should be tumbling wildly out of control,
especially in the post-aphelion period of time on their approach to the Sun.

\section{Augmenting the Database of the SOHO\\Kreutz Populations} 

The revised set of 220 select SOHO sungrazers proved a useful tool for examining
the orbital distribution of the Kreutz system, but there was need to augment the
database in order to remove the deliberate bias against some populations and
thereby to estimate the actual abundances, of Populations~I and II in particular.
This process has three potential steps:\\[-0.25cm]

(a) Expand the population membership classification to all SOHO Kreutz sungrazers,
for which Marsden's gravitational orbits are available and which were imaged
exclusively by the C2 coronagraph (with at least five astrometric observations).
This step should increase primarily the number of members of Populations~I and
Pe, whose presence among the 220 select sungrazers was limited to objects with
at least 12~astrometric observations; and to a lesser extent, the numbers of
members of Populations~\mbox{Pre-I} and Ia, limited to objects with at
least 11 and 10~astrometric observations, respectively.  This step is demonstrated
below to increase the size of the set of 220 select sungrazers by a factor of
more than three.\\[-0.25cm]

(b) Further expand the classification to the SOHO Kreutz sungrazers with Marsden's
orbits, imaged by {\it both\/} the C2 and C3 coronagraphs, by separating for each
object the C2 astrometry from the C3 astrometry and computing a Marsden-like
gravitational orbit only from the former (if the number of observations is at
least five).  This should increase the size of the set made up of the objects
under (a) by probably less than a factor of two.\\[-0.25cm]

(c) Tap the SOHO Kreutz sungrazers' C2 database [$\:\!$i.e., the objects
under (a) and (b)] beyond the final entry of C/2010~M2 in Marsden's set of
1500+~Kreutz objects.  This should increase the size of the set under (b) by
more than a factor of two (as of 2022).\\[-0.25cm]

In the following I get involved with the first step, described in point (a).  A
potentially difficult task is to recognize the SOHO Kreutz sungrazers whose
gravitational orbits computed by Marsden are based exclusively on the C2
astrometric observations.  To achieve this, I employ six sources of data:

(1) A total of 1470 sets of orbital elements for the SOHO comets in Marsden \&
Williams' (2008) {\it Catalogue of Cometary Orbits\/}, starting with C/1996~A2
and ending with C/2008~J16.  This is my primary source of information; it is
complete up to May 14, 2008.

(2) A total of 57 {\it Minor Planet Circulars\/} (MPC), listing orbits for 452
SOHO (and STEREO) comets, starting with C/2008~K1 and ending with C/2010~M2.
The astrometry precedes each orbit in the same batch of MPCs (there are a few
exceptions to this rule).

(3) A total of{\vspace{-0.04cm}} 103 {\it Minor Planet Electronic Circulars\/}
(MPEC),\footnote{See {\tt https://minorplanetcenter.net/mpec/RecentMPECs.html.}}
providing the orbits and astrometry for the same SOHO (and STEREO)
comets.  The difference between the MPCs and MPECs is that starting
with MPEC~2005-E87 in mid-March 2005, the MPECs have explicitly been
discriminating between C2 and C3 astrometric observations.
There has been no such feature in MPC format.

(4) Circulars of the International Astronomical Union (IAUC)\footnote{See
{\tt https://www.cbat.eps.harvard.edu.}} announcing the SOHO (and STEREO) comet
discoveries.  They usually, but not always, specify whether the object was
discovered in C2 or C3 images and later also observed with the other coronagraph.
The IAUCs with discovery announcements of the SOHO Kreutz sungrazers issued before
2000 January~1 are provided in Table~6, showing for each object the respective
MPEC with the astrometry and orbital elements and the MPC with the orbit.

(5) British Astronomical Association Comet Section (BAA)\footnote{See {\tt
https://people.ast.cam.ac.uk/$\sim$jds.}} offers a list of SOHO comets discovered
each year, starting with 2000.  It lists the coronagraph with which the comet was
{\it discovered\/}:\ C2 is inconclusive because a subsequent detection in C3 is
not a priori ruled out; on the other hand, a discovery in C3 eliminates exclusive
imaging with C2.

(6) National Research Laboratory's Sungrazer Project (NRL)\footnote{See
{\tt https://sungrazer.nrl.navy.mil.}} lists the Kreutz sungrazers by year,
column~4 of the table identifying the {\it Telescope\/}.  By clicking on the
perihelion distance in column~5, a link to the relevant MPEC becomes activated.
The lists of the Kreutz objects are grossly incomplete in 2002 and 2003, with
nearly 40 and about 70~objects missed, respectively.

\begin{table}
\vspace{-2.84cm}
\hspace{0.15cm}
\centerline{
\scalebox{0.913}{
\includegraphics{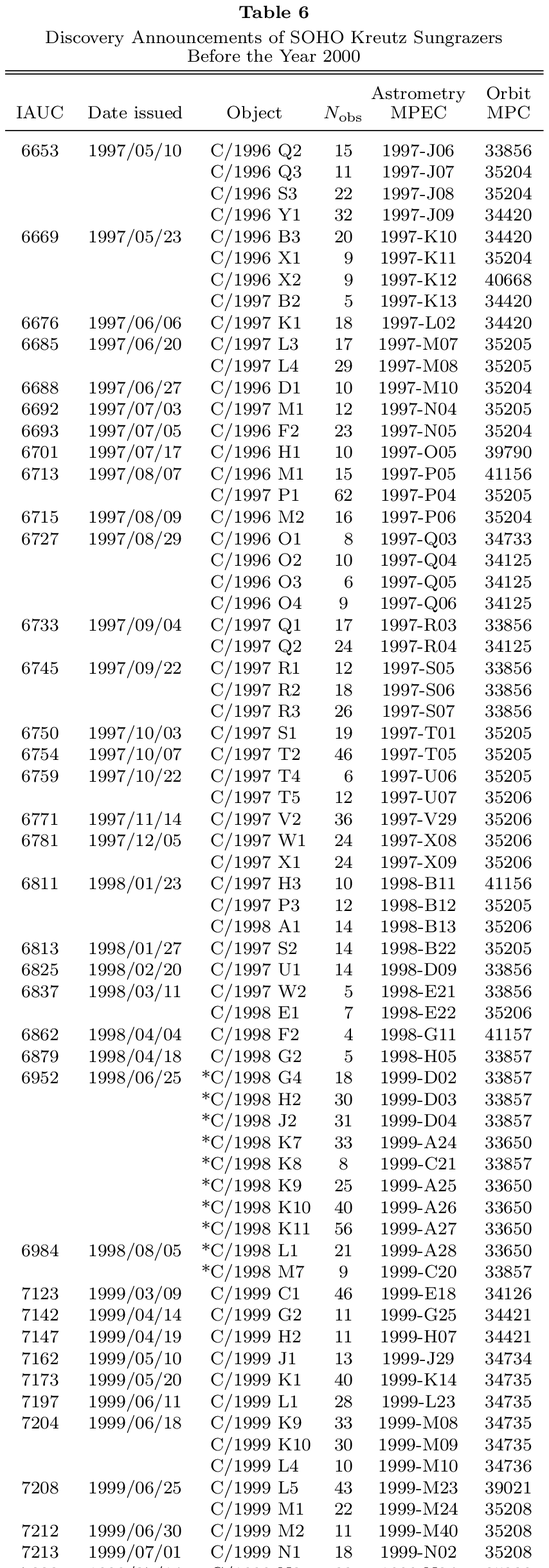}}} 
\vspace{0cm}
\end{table}

Before I continue with the description of my work, I should interject comments
on two issues that the reader should be aware of.  First, the method of
reduction of measured positions of the SOHO comets underwent major changes
after mid-1998.  It appears that C/1998~G2 was the last SOHO comet, whose
positions were reduced the old way (by converting measured polar coordinates
into right ascension and declination; see MPEC 1998-H05).  Starting in
January 1999 at the latest, the equatorial coordinates were determined by
the method of plate constants.  An early notification of this change came
on IAUC~6952, issued on 1998 June~25; it stated that for the nine newly
discovered SOHO comets ``orbit computations are deferred until more definitive
astrometry can be done.''  The eight Kreutz sungrazers among the nine objects,
marked on IAUC~6952 with X/, as well as two more on IAUC~6984, are identified
in Table~6 with asterisks.  From C/1998~G4 on, the astrometry obtained by
applying the new method and the derived orbital elements are presented on
the given MPEC, the orbit also on the MPC.  For C/1998~G2 and all entries
preceding it in Table~6, the old orbits (on the MPEC) were subsequently
recalculated, using the new astrometry (they are on MPCs; a few also on
MPECs are not shown).  The old and new orbital elements for C/1998~G2 are
compared here in Table~7.

The archival discoveries are the second issue I was confronted with.  There
are three groups of them.  The first group includes the SOHO comets detected
early enough that their computed orbits met the deadline for, and are listed
in, the 17th edition of the Marsden-Williams {\it Catalogue\/}.  The second
group includes the objects whose orbits became available between mid-May 2008
and November 2010; these orbits were published by Marsden in one of the MPECs
(such as 2010-L60, 2010-O33, or 2010-T37) and the MPCs (such as 70810, 71681-82,
or 72848-49) from that period of time.  In the third group are the comets
discovered after Marsden had passed away; no orbits are available for these
objects except in some special cases.  It was the second of the three groups
that had to be checked for potential additions to make the set as complete
as possible.

\begin{table}[t]
\vspace{-4.1cm}
\hspace{5.75cm}
\centerline{
\scalebox{1}{
\includegraphics{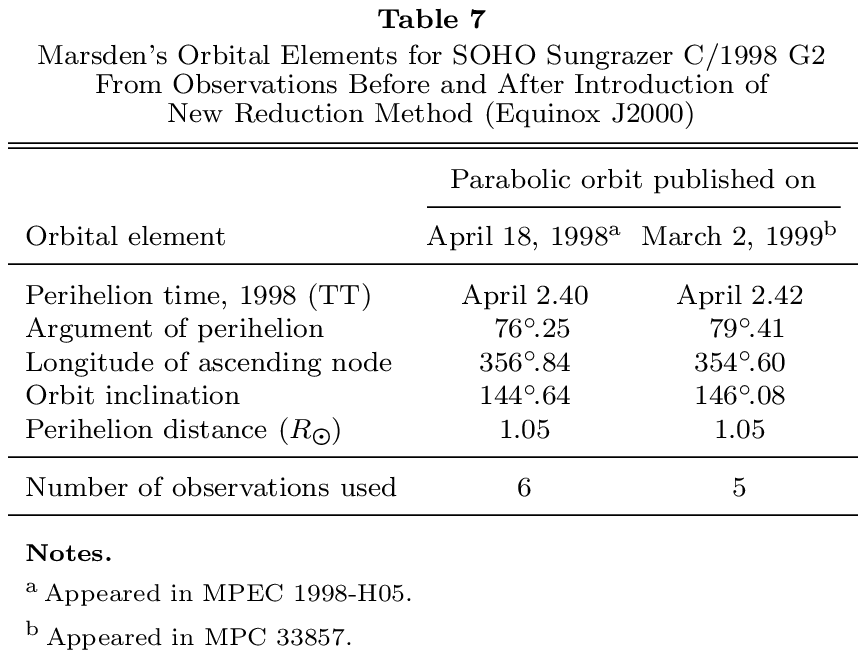}}} 
\vspace{-18.45cm}
\end{table}

To generate a list of SOHO Kreutz sungrazers with Marsden's orbit, imaged
exclusively with the C2 coronagraph at least
five times, I consulted all six sources of data listed above.  In an
overwhelming number of instances, they consistently referred to the
coronagraph(s) involved.  However, before 2005 (see source 3) the task
became challenging in a limited number of cases, when two or more sources
offered conflicting information and it was necessary to decide, which source
is the most likely to be correct.  I do not provide detailed descriptions of
all such instances, but mention one in detail and several additional ones
briefly.

A very confusing situation involved the sungrazers C/2000~M6, C/2000~M7, and
C/2000~M8.  The last of the three is one of the entries removed from the
set of 193~select Kreutz objects (cf.\ Table~2).  It was initially (in
Paper~1) included in the set on the strength of the claim by NRL that C/2000~M8
was observed with C2, while C/2000~M6 and C/2000~M7 with C2 and C3.  More
recently I checked the BAA source, according to which C/2000~M6 was discovered
with C3 while C/2000~M7 and C/2000~M8 with C2.  This information is clearly in
line with NRL for C/2000~M6 and C/2000~M8 and not in conflict for C/2000~M7:\
although the object would of course appear in C3 first, it might have been
overlooked and its detection in C3 only later confirmed.  The doubts arose
after I checked the original report on IAUC~7453, which claims that C/2000~M6
was observed only with C2, while C/2000~M7 and C/2000~M8 with C3 and C2.  This
is inconsistent with both NRL (for C/2000~M6 and C/2000~M8) and BAA (at least
for C/2000~M6).  MPEC 2000-N31 lists 14~observations spanning 0.19~day for
C/2000~M6, 40~observations spanning 0.74~day for C/2000~M7, and 21~observations
spanning 0.38~day for C/2000~M8.  This suggests that the information on the IAU
Circular is almost certainly the correct one as a Kreutz sungrazer is seldom
observed in C2 over a period of time longer than 7~hours.

Several additional contradictions of the same type:\ (i)~C/2002~K6 was claimed
to be discovered in C3 by BAA, but observed only in C2 by NRL.  It was reported
in IAUC~7913 as found in C2 and also seen in C3; (ii)~C/2002~X13 was discovered
in C3 according to BAA, left out by NRL, and found with and visible only in
C2 according to IAUC~8266; (iii)~C/2003~M3 is listed as discovered in C3 by
BAA, but in C2 by IAUC~8327, being ignored by NRL; and (iv)~C/2004~V6 and
C/2004~V7 are particularly confusing:\ BAA says they were discovered in C2 and
C3, respectively, while NRL and IAUC~8455 disagree on their visibility:\ the
first object was seen in C3 and C2 and the second only in C2 according to NRL,
but the other way around according to the IAU Circular; BAA is not in conflict
with the Circular but it is with NRL (in the case of the second object), yet
MPEC~2004-X72 shows the observations of C/2004~V6 and C/2004~V7 covered
0.37~day and 0.08~day, respectively, implying that it was the latter object
that was followed in C2 only; this is in line only with NRL.

I could submit a number of other similar contradictory cases, but the few examples
should convince the reader that none of the sources listing the instrument appears
to be flawless.  Accordingly, one likewise should expect a few problems with the
following tabulated data, which summarize the population membership classification
for all SOHO Kreutz comets with Marsden's gravitational orbits derived exclusively
from the C2 astrometry (with a minimum of five observations).  To be absolutely
certain that all objects satisfying the criteria are in the set and all objects
not conforming to the criteria have been eliminated from it, one would have to
compare the time of each observation with the archive of SOHO imaging times,
a task deemed here prohibitively time consuming to undertake.

The classification of the population membership in Tables~8--11 is based on
the computed true longitude of the ascending node and its comparison with the
populations' boundaries in Table~1.  Since the wings of the major populations,
Population~I in particular, spill across these boundaries, the actual membership
of borderline objects is questionable.  The 753~sungrazers, 48~percent of all
SOHO Kreutz sungrazer orbits that Marsden derived, are chronologically divided
into four groups:\ 202~comets from the years 1997--2001 are in Table~8,
188~comets from 2002--2004 in Table~9, 195~comets from 2005--2007 in Table~10,
and the remaining 168~comets from 2008--2010 in Table~11.  No object from 1996.
The 220~select sungrazers from the limited set are marked with asterisks.

Not listed in the tables are seven objects with extreme orbital properties.
Four probably are not Kreutz sungrazers:\ C/2001~N1, whose inclination is
95$^\circ$; C/2007~A7, whose perihelion distance is 4.1\,{\Rsun}; C/2007~M5,
whose inclination is 154$^\circ$ and perihelion distance is 0.24\,{\Rsun};
and C/2009~Y9, whose inclination is 152$^\circ$ and argument of perihelion
43$^\circ\!$.5.  The remaining three --- C/2007~W5, C/2007~Y6, and C/2008~W12
--- may have the nominal elements within a range acceptable for Kreutz comets,
but imply anomalous{\vspace{-0.085cm}} values for the true longitude of the
ascending node, $\widehat{\Omega}$:\ 21$^\circ\!$.5, 17$^\circ\!$.2, and
31$^\circ\!$.6, respectively.

One objective for expanding the present investigation to cover the large data
set was to generate a histogram of the true longitudes of the ascending node free
from the deliberate bias of the limited sets in order to obtain a good estimate for
the relative abundances of Populations~I and II.  The second objective was to get
potentially new evidence on the other populations discovered in Paper~1 in the
set of 193 sungrazers.

\begin{table*}[t]
\vspace{-3.72cm}
\hspace{0.15cm}
\centerline{
\scalebox{0.915}{
\includegraphics{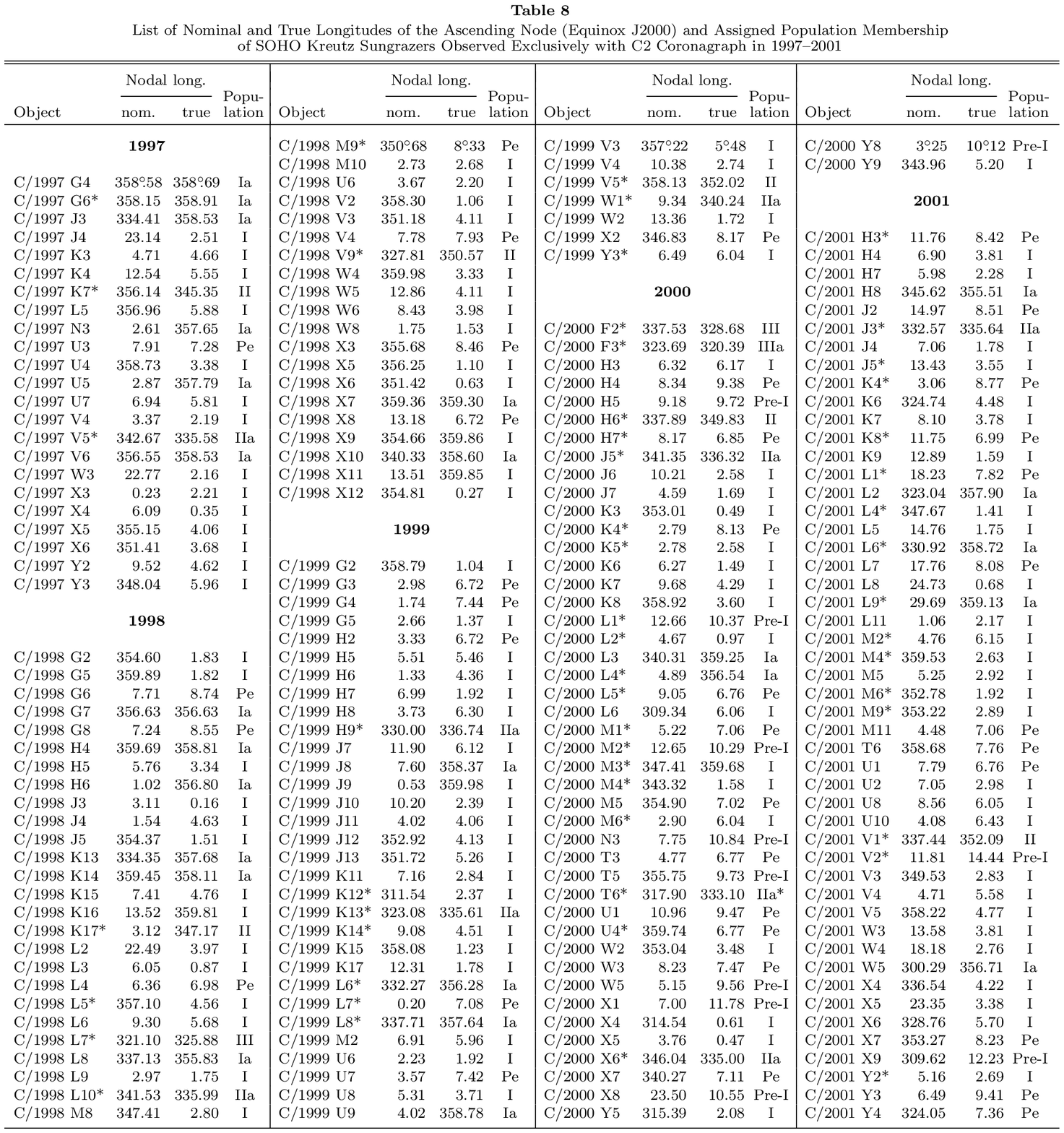}}} 
\vspace{-3.57cm}
\end{table*}

The histogram of the true nodal longitudes for the 753 SOHO Kreutz comets is
displayed in Figure 2.  The significance of introducing the true longitude
is plainly demonstrated by comparing the distribution in this histogram with
that of the {\it nominal\/} nodal longitudes of the same objects, which is
shown to the same scale in \mbox{Figure}~3.  It is noted that the sharp, discrete
peaks and a limited range of the nodal longitudes in Figure~2 are replaced
in Figure~3 with a noisy pseudo-Gaussian distribution, whose flat,
inconspicuous peak near the nodal longitude of the Great March Comet of 1843
is less than one half the size of the peak of Population~I in Figure~2.  Not
counting the side branch Pe, the Population~I-to-II estimated abundance ratio
is at least 14:1 in Figure~2, much higher than in Figure~1.

\begin{table*}[t]
\vspace{-3.72cm}
\hspace{0.15cm}
\centerline{
\scalebox{0.915}{
\includegraphics{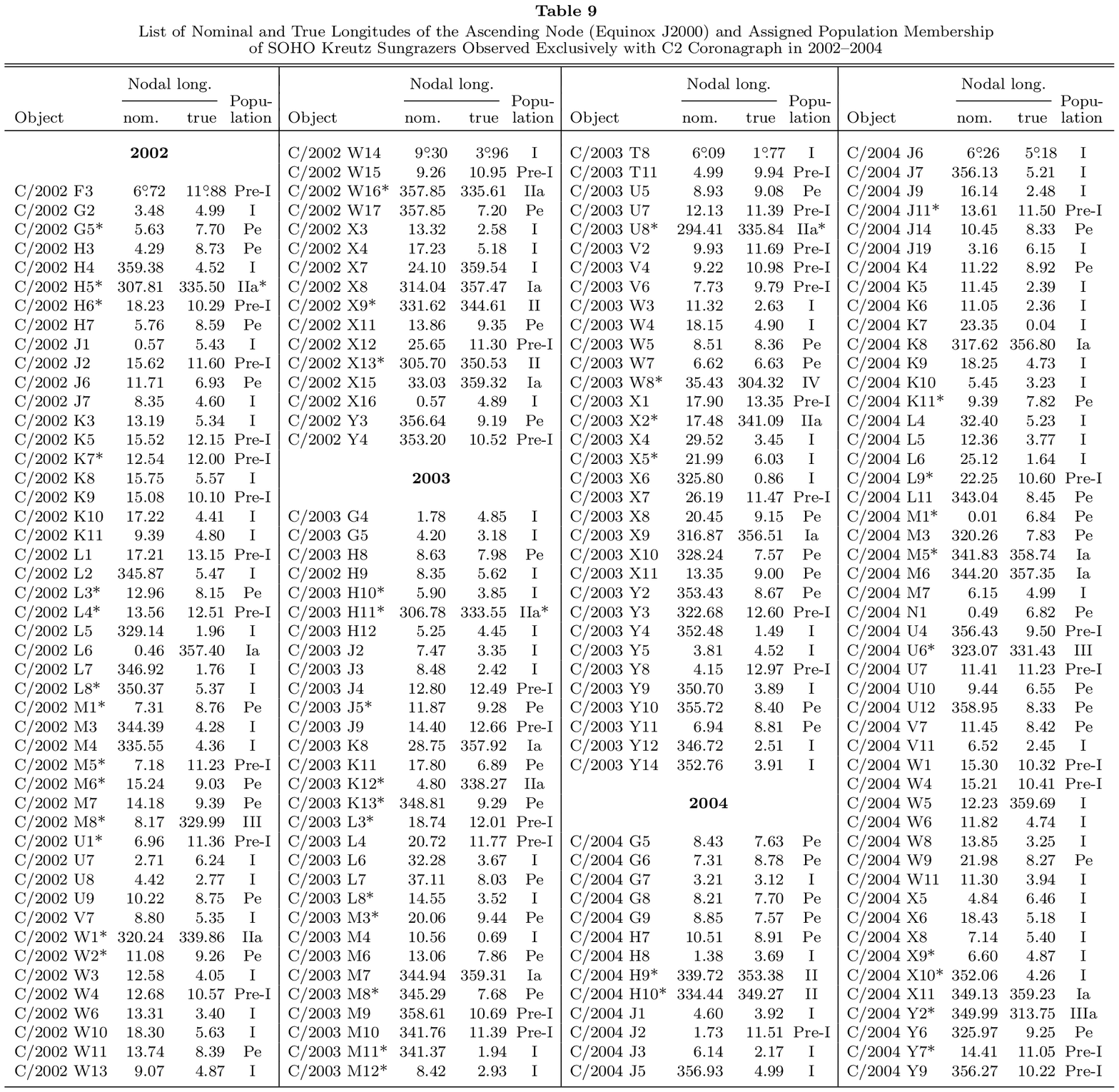}}} 
\vspace{-4.7cm}
\end{table*}

In regard to the second objective, Figure~2 confirms that the bias of the
limited dataset against Population~I and its side branch Pe was helpful in
that it allowed the nearby less prominent Populations~Ia and \mbox{Pre-I}
to stand out.  The dominance of the unrestricted Populations~I and Pe is so
overwhelming that Populations~Ia and \mbox{Pre-I} become almost completely
absorbed by the wings of the neighboring populations and protrude only
as minor bulges along the slopes of the histogram in Figure~2.  The second
peak of Population~I not only confirms its existence established from the
limited set, but now surpasses the peak of Population~Pe{\vspace{-0.04cm}}
in size and moves from \mbox{$\widehat{\Omega} = 1^\circ$} to 2$^\circ$.  The
large number of dwarf sungrazers in this part of Population~I is bound to imply
the existence of a bright, naked-eye sungrazer, which somewhat inexplicably
has as yet remained undetected.

The other populations from the limited dataset appear to remain in Figure~2
essentially unchanged, with decreasing prominence from IIa through III and
IIIa to IV.  I see no evidence for any additional populations.  The anomalous
values of the true nodal longitude for the three eliminated potential Kreutz
sungrazers were widely scattered and it would be preposterous to take them
for some sort of seeds of new populations.\\[-0.5cm]

\begin{table*}[t]
\vspace{-3.72cm}
\hspace{0.15cm}
\centerline{
\scalebox{0.915}{
\includegraphics{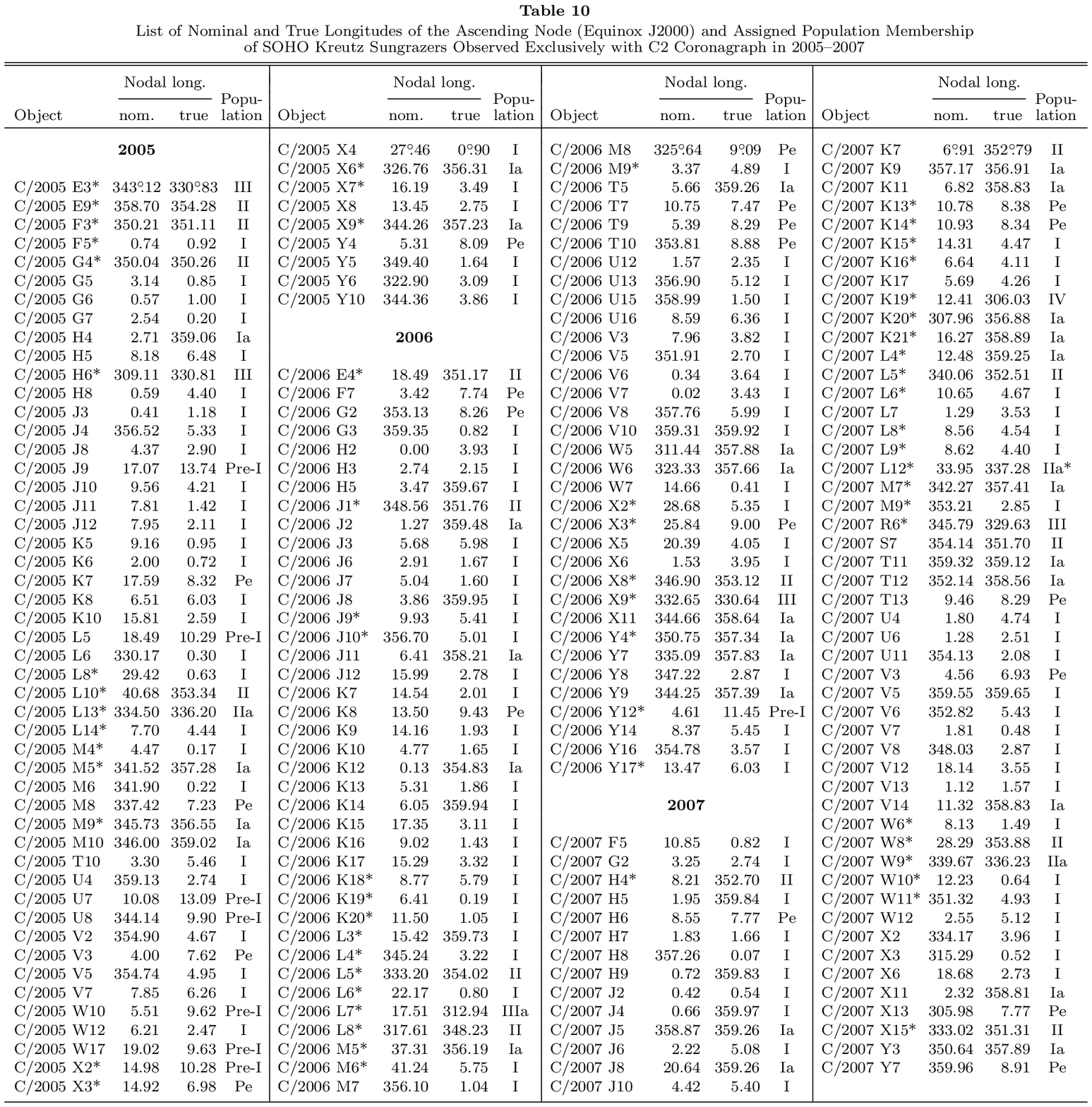}}} 
\vspace{-4.22cm} 
\end{table*}

\section{Possible Explanation of the Poor Showing of Population~II Relative
to Population~I Among the SOHO Sungrazers} 
In Paper 2 I developed a simple model of Population~I among the SOHO dwarf
sungrazers, which showed the range of their orbital dimensions to be enormously
protracted.  Because these objects do not survive their first return to
perihelion, their life span does not exceed their orbital period.

\begin{table*}[t]
\vspace{-3.72cm}
\hspace{0.15cm}
\centerline{
\scalebox{0.915}{
\includegraphics{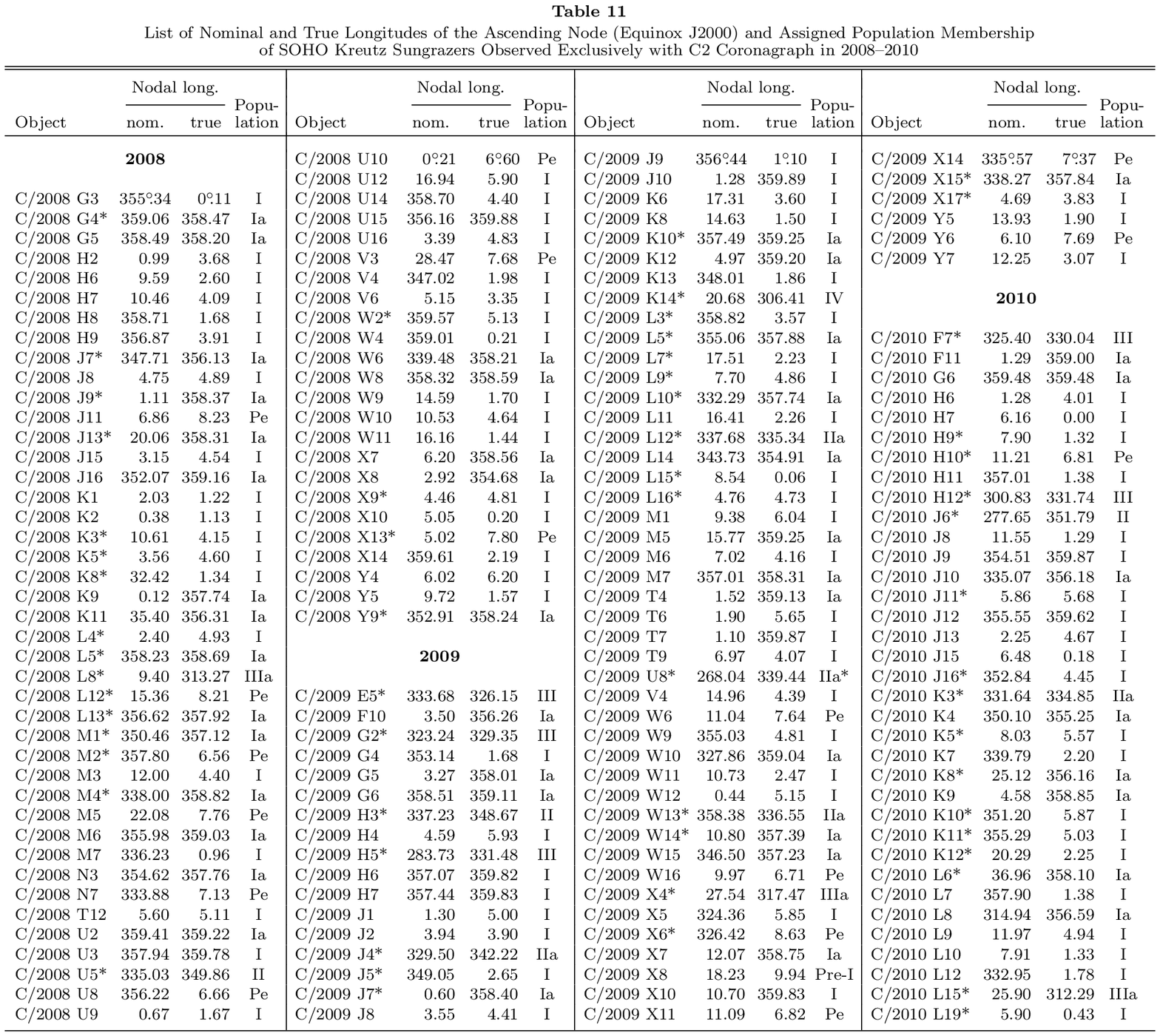}}} 
\vspace{-6.4cm}
\end{table*}

I employed a mechanism of spontaneous separation of fragments from the surface
of a sizable parent comet nucleus.  At the moment of its separation, a fragment
shared the parent's orbital velocity but its heliocentric distance differed,
in general, from the heliocentric distance of the parent's center of mass.
These two conditions suffice to warrant that the fragment ended up
in an orbit different, potentially very different, from the parent's.  The
orbital period of the fragment was shorter than the comet's for a separation
from the sunward hemisphere; it was longer for a separation from the other
hemisphere.  In Paper~2 I showed that the relation between the fragment's
orbital period, $P_{\rm frg}$, and the parent comet's orbital period, $P_{\rm
par}$, is given by an expression
\begin{equation}
P_{\rm frg} = P_{\rm par} \left[ 1 - \frac{2 \Delta U}{r_{\rm frg}^2}
  P_{\rm par}^{\frac{2}{3}} \right]^{-\frac{3}{2}}\!\!, 
\end{equation}
where $\Delta U$ is the difference between the heliocentric distances (in AU)
in the sense ``fragment minus parent's center of mass'' and $r_{\rm frg}$ is the
heliocentric distance of the fragment's separation (also in AU).{\vspace{-0.09cm}}
Note that when the orbital period is in years, $P_{\rm par}^{2/3}$ numerically
equals the parent's semimajor axis in AU and the second term in the brackets
becomes dimensionless, as expected.

Applying this formula to the SOHO sungrazers of Population~I, whose parent was,
in the framework of the contact-binary model, the Great Comet of 1106 (X/1106 C1),
I assumed that an ensemble of fragments separated from the entire surface of
the comet at perihelion.  For the sake of argument I adopted that the nucleus
extended for 50~km along the radius vector, so that $\Delta U$ varied from $-$25~km
to +25~km{\vspace{-0.06cm}} (or from \mbox{$-\frac{1}{6}\times 10^{-6}$\,AU} to
\mbox{$+\frac{1}{6}\times 10^{-6}$\,AU}).  A SOHO-like swarm of dwarf comets
was returning to perihelion starting in the late 14th century (the fragments
at \mbox{$\Delta U \simeq -25$}~km) and it will continue for nearly 80,000~yr
(fragments at \mbox{$\Delta U \simeq +25$}~km), even though at dramatically
lower rates.  In about 3000~yr from now the rate was in Paper~2 predicted
to drop about ten times from today's.

\begin{figure*}[t]
\vspace{-6cm}
\hspace{-0.16cm}
\centerline{
\scalebox{0.6525}{
\includegraphics{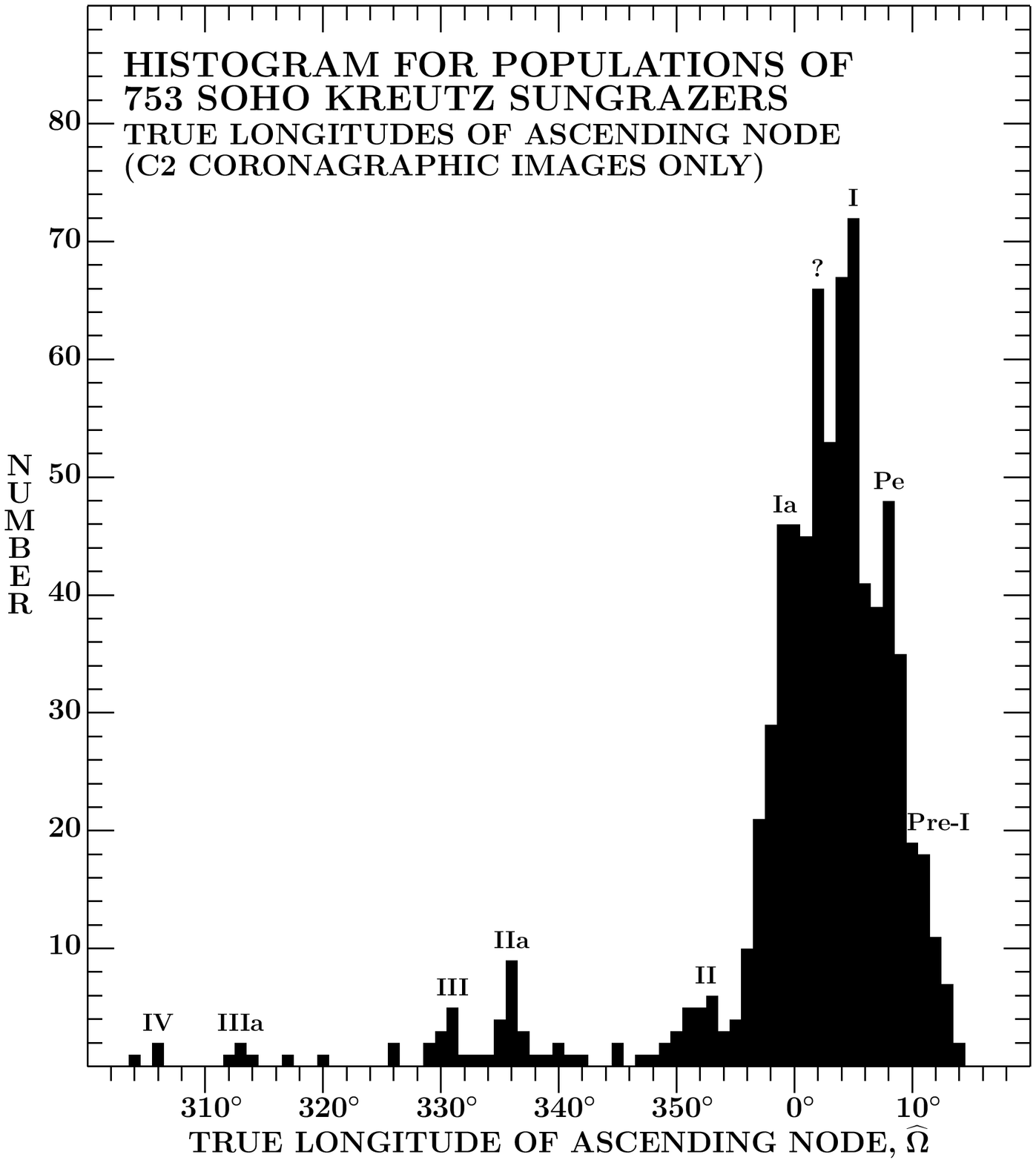}}} 
\vspace{-0.47cm}
\caption{Histogram of the true longitudes of the ascending node for
the complete set of 753 SOHO Kreutz sungrazers from~1997~to~2010,
whose orbits were derived by Marsden exclusively from the
astrometry of C2 coronagraphic images (a minimum of five
positions).  The nine populations are marked.  Population~Pe
is considered a side branch of Population~I.  The two peaks of
Population~I now dominate the distribution of the nodal
longitudes.  On the other hand, Populations~Ia and \mbox{Pre-I}
are hidden by the swollen wings of, respectively, Populations~I
and Pe.  Compared to Figure~1, Populations~II through IV are
changed little or not at all.  The Population~I-to-II abundance
ratio is estimated at not less than 14:1.{\vspace{0.7cm}}}
%
\vspace{-10.45cm}
\hspace{-0.16cm}
\centerline{
\scalebox{0.87}{
\includegraphics{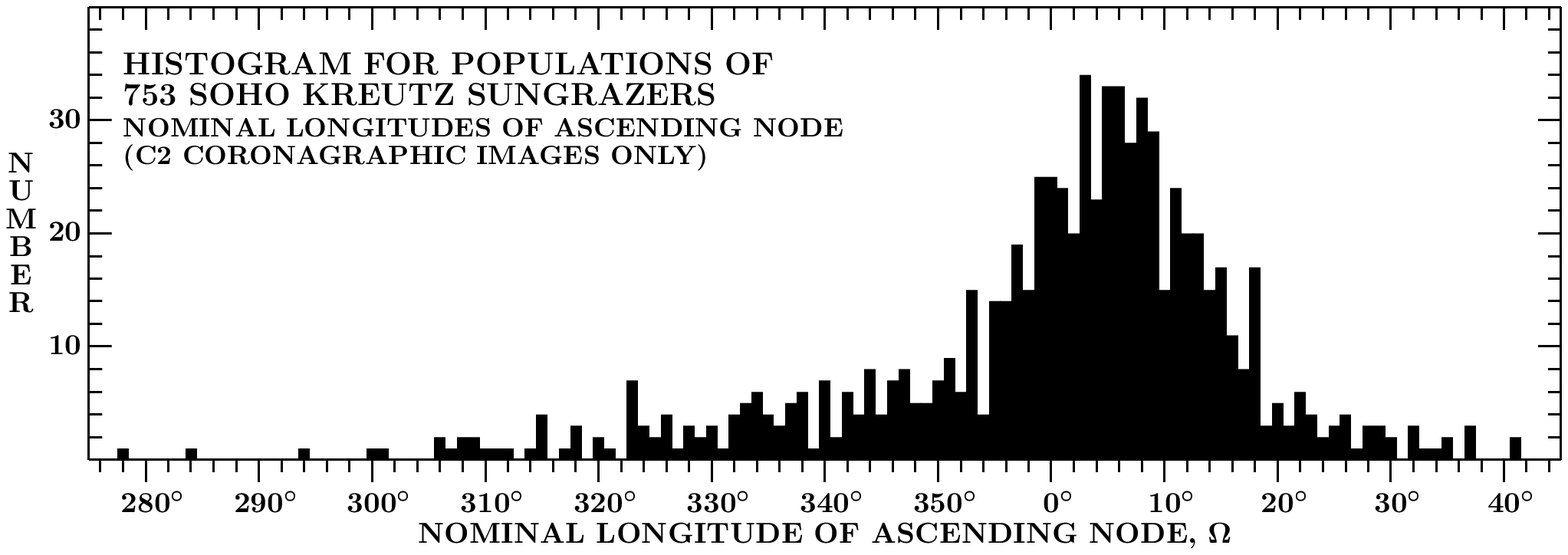}}} 
\vspace{-8.79cm}
\caption{Histogram of the nominal longitudes of the ascending
node, derived by Marsden for the same set of 753 SOHO Kreutz
sungrazers, as in Figure~2.  Another single entry is at
\mbox{$\Omega = 268^\circ$}.  The distribution of the nodal
longitudes looks noisy and much broader, an effect of the
normal component of the sublimation-driven nongravitational
acceleration.  The peak is poorly defined, but rather close
to the nodal longitude of the Great March Comet of 1843, and
its size is barely one half the size of the most prominent
peak of Population~I in Figure~2.{\vspace{0.7cm}}}
\end{figure*}     

Examining the production of SOHO comets in Population~II, I now assume that
their birth coincided with the splitting of the Chinese comet of 1138 into
the Great September Comet of 1882 (C/1882~R1) and comet Ikeya-Seki (C/1965~S1).
Sekanina \& Kracht (2022) derived for the 1138 comet a perihelion distance of
1.73\,{\Rsun} and argued that its breakup into the 1882 and 1965 sungrazers
took place nearly 2$\frac{1}{2}$~hr after perihelion at a heliocentric
distance of 4.15\,{\Rsun} or 0.0193~AU.  For the diameter of the 1138 parent
the authors adopted a value of 60~km, so that $\Delta U$ varied in the
range from $-$30~km to +30~km.  The orbital period of the 1138 comet was
\mbox{$P_{\rm par} = 754$}~yr (Sekanina \& Kracht 2022), so that Equation~(12)
gives at the two limits of $\Delta U$ the orbital periods of 663~yr and 867~yr.
This simple model predicts that the fragments should have been returning to
perihelion over a period of about two centuries between AD~1801 and 2005.
The swarm of Population~II is thus predicted to cease in the middle of the
investigated SOHO database.  Given the uncertainties of the estimate, one
is justified to make a preliminary conclusion that a possible reason for
the poor showing of the Population~II SOHO comets is the physical termination
of the stream because of the dynamical constraints.  Future observations will
show whether this conclusion is correct.

\section{Closing Comments} 
This study exploits the opportunity offered by the fixed slope of the
relationship between the nominal longitude of the ascending node and the
nominal latitude of perihelion among the SOHO Kreutz sungrazers.  This
feature puts forward an option of portraying the distribution of their
population membership in a histogram of the {\it true\/} longitude of the
ascending node, a method implemented in this paper for the first time.
The database is limited to the dwarf sungrazers observed (on at least
five occasions) exclusively with the C2 coronagraph on board the SOHO
spacecraft, whose imager has a small enough pixel size to allow, unlike
the C3 coronagraph, fairly accurate astrometry.  Even though the Kreutz
sungrazers detected in C2 as well as in at least one of the COR2
coronagraphs on board the two spacecraft of the {\it Solar Terrestrial
Relations Observatory\/} (STEREO) --- one is no longer operational at
present --- have orbital elements of higher quality than those imaged with
C2 alone, too few such objects are available for a meaningful statistical study.  

By introducing the {\it true\/} longitude{\vspace{-0.055cm}} of the ascending
node, $\widehat{\Omega}$, equaling its nominal value that fits the Kreutz
system's standard latitude of perihelion, each sungrazer is assigned a single
parameter, which approximately corrects the motion for effects of the normal
component of the sublimation-driven nongravitational acceleration.  The
histogram for a revised set of 220~select Kreutz sungrazers displays the
nine populations, previously detected in the plots of the two nominal
quantities for a smaller set of 193~sungrazers.  In the histogram, all nine
populations are apparent, but{\vspace{-0.055cm}} Population~I shows two
peaks located nearly symmetrically in $\widehat{\Omega}$ relative to the
nodal longitude of the Great March Comet of 1843.

The limited set of 220 select SOHO Kreutz sungrazers is biased deliberately
to reduce the overwhelming contribution from Populations~I and Pe and thereby
to allow the less prominent Populations~Ia and \mbox{Pre-I} to stand out.  The
limited set is also used to single out groups of the Kreutz sungrazers related
to one another by sharing essentially the same value of an orbital element.
However, because of the lack of algorithm, the nominal perihelion distance
could not be converted to a {\it true\/} perihelion distance, so this
element provides no information on the fragmentation process that gradually
modifies the size distribution of SOHO-like fragments in a stream with
time.  On the other hand, the true longitude of the ascending node and the
perihelion time are both free from major effects of the nongravitational
forces, which makes the statistical averages of their quadratic differences
of diagnostic value in constraining the preferred locations of fragmentation
events along the orbit.  Fragmentation appears to have proceeded along the
entire orbit, both before and after aphelion.  The separation velocities,
although difficult to determine, may have been as high as $\sim$1~m~s$^{-1}$.
If rotational in nature, they would suggest rapid, out-of-control tumbling
of the fragments.  

Next, I augmented the revised set of select objects to {\it all\/} SOHO
Kreutz sungrazers, for which Marsden determined an orbit and which were
imaged exclusively with the C2 coronagraph (at least on five occasions)
to obtain a more representative abundance ratio between Populations~I and
II, as well as to detect potential changes in the other populations.  I
described pitfalls encountered in an effort to recognize the C2-only objects
and determined that in the augmented set of 753 SOHO sungrazers Population~I
overwhelmed Population~II at a ratio of at least 14:1.  Of the other
populations, Pe remained clearly visible in the histogram in Figure~2,
but Ia and \mbox{Pre-I} were not prominent enough to avoid being essentially
overrun by the wings of Populations~I and Pe, respectively.  As a result,
Populations~Ia and \mbox{Pre-I} showed up merely as minor bulges on the
slopes of the histogram.  On the other hand, Populations~II through IV
changed very little or not at all.

Clearly of interest was to compare the histograms of the {\it true\/} and {\it
nominal\/} longitudes of the ascending node.  The difference between the two
was shown to be enormous, the relatively narrow range of the nodal longitudes
and the discrete population peaks in the former contrasting with a much
wider range and a noisy, pseudo-Gaussian distribution of the longitudes in the
latter.  The magnitude of either of the two prominent peaks of Population~I's
true nodal longitudes is in Figure~2 about twice the magnitude of the maximum
of the flat distribution of nominal nodal longitudes in Figure~3.

The present investigation confirms the existence of the nine populations among
the SOHO Kreutz sungrazers discovered in Paper~1.  The higher peak of Population~I
is probably related directly to the Great March Comet of 1843, while the lower
peak at a (true) nodal longitude of 1--2$^\circ$ is likely to refer to another
side branch similar to Pe, except that, unlike C/1963~R1, its naked-eye member
is still to be discovered.  Although the high abundance ratio of Population~I
to Population~II could be affected by temporal issues in fragment release, the
increasingly dramatic drop in the number of detected sungrazers on the side
of Population~II is in line with the fragmentation scenario proposed in Paper~1
and plainly visible from Figure~6 of Paper~2:\ following the primary breakup
of the progenitor, ever smaller amounts of the surviving mass of Lobe~II were
available, as one proceeds from Fragment~II to Fragments~IIa*, III, IIIa, and
IV, where the first phase of the advancing process of cascading fragmentation
appears to finally come to an end.  Populations~II, IIa, III, IIIa, and IV are
seen in Figures~1 and 2 to rather faithfully emulate the early trends in the
proposed evolution of the Kreutz system, even though mass-wise on scales orders
of magnitude smaller.\\

This research was carried out at the Jet Propulsion Laboratory, California
Institute of Technology, under contract with the National Aeronautics and
Space~Administration.{\pagebreak}

\begin{center}
{\footnotesize REFERENCES}
\end{center}

\vspace{-0.25cm}
\begin{description}
{\footnotesize
%
%
\item[\hspace{-0.3cm}]
Battams, K., \& Knight, M. M. 2017, Phil.\ Trans.\ Roy.\ Soc.\ A, 375,{\linebreak}
 {\hspace*{-0.6cm}}2097
\\[-0.57cm] 
%
%
%
%
%
%
%
%
%
%
%
\item[\hspace{-0.3cm}]
Kreutz, H. 1901, Astron. Abhandl., 1, 1
\\[-0.57cm]
%
%
%
\item[\hspace{-0.3cm}]
Marsden, B. G. 1967, AJ, 72, 1170
\\[-0.57cm]
\item[\hspace{-0.3cm}]
Marsden, B. G. 1989, AJ, 98, 2306
\\[-0.57cm]
\item[\hspace{-0.3cm}]
Marsden, B. G. 1990, in Asteroids, Comets, Meteors III, ed.\ C.-I.{\linebreak}
 {\hspace*{-0.6cm}}Lagerkvist, H.\ Rickman, \& B.\ A.\ Lindblad (Uppsala:\
 Univer-{\linebreak}
 {\hspace*{-0.6cm}}sitet), 393
\vspace{0.4cm}
\item[\hspace{-0.3cm}]
Marsden, B. G. 2005, Annu. Rev. Astron. Astrophys., 43, 75
\\[-0.57cm]
\item[\hspace{-0.3cm}]
Marsden, B. G., \& Williams, G. V. 2008, Catalogue of Cometary{\linebreak}
 {\hspace*{-0.6cm}}Orbits 2008, 17th ed. Cambridge, MA: Minor Planet Center/{\linebreak}
 {\hspace*{-0.6cm}}Central Bureau for Astronomical Telegrams, 195pp
\\[-0.57cm]
%
%
%
%
%
%
%
\item[\hspace{-0.3cm}]
Sekanina, Z. 2000, ApJ, 542, L147
\\[-0.57cm]
%
%
\item[\hspace{-0.3cm}]
Sekanina, Z. 2021, eprint arXiv:2109.01297 (Paper 1)
\\[-0.57cm]
%
%
%
\item[\hspace{-0.3cm}]
Sekanina, Z. 2022, eprint arXiv:2211.03271 (Paper 2)
\\[-0.57cm]
%
%
%
%
\item[\hspace{-0.3cm}]
Sekanina, Z., \& Chodas, P. W. 2012, ApJ, 757, 127 (33pp)
\\[-0.57cm]
\item[\hspace{-0.3cm}]
Sekanina, Z., \& Kracht, R. 2015, ApJ, 801, 135 (19pp)
\\[-0.67cm]
\item[\hspace{-0.3cm}]
Sekanina, Z., \& Kracht, R. 2022, eprint arXiv:2206.10827}
%
%
%
%
\vspace{0.05cm}
\end{description}
\end{document}